# Remarks on the mathematical origin of wave mechanics and consequences for a quantum mechanics in a gravitational field[*]

Mayeul Arminjon

Laboratoire 3S, Institut de Mécanique de Grenoble,
B.P. 53, 38041 Grenoble cedex 9, France.

**Abstract.** According to Schrödinger's ideas, classical dynamics of point particles should correspond to the « geometrical optics » limit of a linear wave equation, in the same way as ray optics is the limit of wave optics. It is shown that, using notions of modern wave theory, the « geometrical optics » analogy leads to the correspondence between a classical Hamiltonian $H$ and a « quantum » wave equation in a natural and general way. In particular, the correspondence is unambiguous also in the case where $H$ contains mixed terms involving momentum and position. In the line of Schrödinger's ideas, it is also attempted to justify the occurrence, in QM, of eigenvalues problems, not merely for energy, but also for momentum. It is shown that the wave functions of pure momentum states can be defined in a physically more satisfying way than by assuming plane waves. In the case of a spatially uniform force field, such momentum states have a singularity and move undeformed according to Newton's second law. The mentioned unambiguous correspondence allows to uniquely extend the Klein-Gordon relativistic wave equation to the case where a constant gravitational field is present. It is argued that Schrödinger's wave mechanics can be extended to the case with a variable gravitational field only if one accepts that the wave equation is a preferred-frame one. From this viewpoint, generally-covariant extensions of the wave equations of QM seem rather formal. Finally, it is conjectured that there is no need for a quantum gravity.

## 1. Introduction and statement of the main results

De Broglie found that the analogy between the variational principles which arise in geometrical optics and in Hamiltonian dynamics had its origin in a wave-particle duality valid for material particles as well as for electromagnetic radiation. Elaborating on de Broglie's ideas, Schrödinger emphasized that geometrical optics is merely an approximation to wave optics and, starting from the Hamilton-Jacobi equation of classical mechanics, he proposed his famous wave equation. For Schrödinger, the « matter waves » had to be generally more real than the particle aspect. Schrödinger's way to his equation was rather intuitive – although, in the application of this equation, he mastered advanced mathematics of his time. In today's quantum mechanics, the correspondence between a classical Hamiltonian and a « quantum » wave equation, as well as the rules of measurement, are introduced in a formal, axiomatic way (perhaps after having motivated this axiomatic by the examination of elementary examples). Moreover, the wave function is interpreted, not as representing a real wave, but instead as a tool to compute probabilities. While the quantum axiomatic works extremely well in the innumerable usual applications of quantum mechanics, it leads to ambiguities when one tries to extend it to a situation with curved space-time – which is needed if one wishes to have a quantum mechanics in a gravitational field. Furthermore, the experimentally confirmed predictions of quantum mechanics, in situations of the Einstein-Podolsky-Rosen type, are difficult to interpret in the framework of physical realism. A realistic interpretation might be made easier if one would abandon the orthodox notion that the quantum objects are more likely to consist of point particles, and if one would come back, instead, to Schrödinger's concept according to which his extended waves describe an essential element of the physical reality of a quantum object. If, furthermore, one would partly relax the relativistic interdiction to pass beyond the velocity of light, the difficulties of interpretation might be even smaller [16]. The Lorentz-Poincaré version of special relativity gives support to this attitude, because it « desacralizes » the role of *c*. Due to relativistic mechanics, *c* remains an upper limit for the velocity of *classical particles*, but the fact that an ether is preserved in this version [12] allows to interpret the limiting velocity in an intuitive and less absolute way. If one thinks of material particles as of topologically structured flows in a fluid ether, the limit *c* applies to the organized flow as a whole, rather like the sound velocity [13], and does not forbid that parts of the flow may be superluminal – which is indeed the case in the slightly different context of purely electromagnetic particles [19].

The present work is an attempt to understand Schrödinger's wave mechanics in the light of the modern theory of (classical) waves, to give some (incomplete) support to the possibility of a realistic interpretation of his waves, and to draw conclusions for a possible extension of quantum mechanics to a situation with a gravitational field.

First, the correspondence between a linear wave equation and a classical Hamiltonian (algebraic in **p**) will be analysed in the framework of the theory of dispersion relations in a linear medium. It will be shown that this correspondence is fully justified under the assumption that the classical Hamiltonian should describe the trajectories corresponding to the « geometrical optics » limit of the considered wave equation. Thus, the wave equation deduced from a classical Hamiltonian is *a priori* no more « quantum » than is the d'Alembert equation for the classical electromagnetic potential. This correspondence is *one-to-one* also in the general case where the Hamiltonian contains mixed terms involving momentum **p** and position **x** (and time). This allows to use it also in the case with a gravitational field, the latter making the (extended) configuration space heterogeneous.

In the second part of the paper, it will be attempted to understand, from the point of view of wave mechanics, how eigenvalue problems do occur in quantum mechanics. In preparation for the gravitational context, it will be briefly recalled how stationary states are related to *energy states*, *i.e.* to eigenvalue problems for the spatial Hamiltonian operator. Then, the case of momentum states will be discussed in some detail. If one follows the principles of *wave mechanics*, the most essential property of a « *pure momentum state* » is that the group velocity should be spatially uniform. This property is true for a plane wave in the case of the usual Schrödinger equation, of course, but it is shared with a plenty of different wave functions. There are additional properties which should be satisfied by the wave function of a « pure momentum state » : it should obey the time-dependent Schrödinger equation, and it also should satisfy the dispersion relation expressing the frequency as a function of the wave vector. Furthermore, if one wishes that such wave function may represent a single quantum object, it should be spatially limited in the minimum sense that the amplitude should vanish at infinity – a property obviously wrong for plane waves. It will be shown that these properties, taken together, lead to the unavoidable conclusion that:

(i) pure momentum states can exist only in so far as the spatial variation of the force field is negligible, *i.e.*, the potential must be linear in **x**;

(ii) in that case, they do exist (thus for whatever time-dependence of the force field), and they are characterized by a wave which undergoes a mere translation with time, thus *no deformation and, in particular, no spreading out at all;*

(iii) they *must* have a *singularity*, whose motion obeys *Newton's second law*.

The third part of the paper is devoted to the extension of wave mechanics in the presence of a gravitational field, as characterized by a curved space-time metric. It will be shown that, with a curved metric, Schrödinger's wave mechanics is inherently frame-dependent, hence a consistent wave mechanics in a gravitational field implies that a *physically preferred reference frame* must exist. This is, of course, incompatible with general relativity, but it is fully consistent with the « ether theory of gravitation » studied by the author. Thus, although one may propose gravitational extensions of quantum-mechanical wave equations without assuming any preferred reference frame, this seems rather formal from the viewpoint of Schrödinger's wave mechanics. In particular, it will be proved that a classical Hamiltonian exists for the motion of a test particle moving « freely » in a *static* gravitational field; that the wave-mechanical correspondence leads to unambiguously associate with this Hamiltonian a wave equation, which is thus *the extension of the free Klein-Gordon equation;* and that this wave equation, which also makes sense for a general gravitational field, *is definitely a frame-dependent equation* (in contrast to the obvious manifestly covariant extension of the free Klein-Gordon equation). The stationary states of this equation are defined: in the static case, they are solutions of an eigenvalue problem for a spatial operator. In the general case, one still may define stationary states, but they they do not seem to correspond to a spatial eigenvalue problem.

Moreover, it will be shown that the « wave-mechanically correct » extension of the free Klein-Gordon equation has the property that the associated energy-momentum tensor **T** *does not satisfy* $T^{\mu\nu}{}_{;\nu} = 0$, even in the static case. The latter equation is the dynamical equation in general relativity and other metric theories, and also, for the static case, in the preferred-frame theory studied by the author. Hence, it seems that, in the presence of a gravitational field, one has to make an unpleasant choice between wave mechanics and the most general form of classical dynamics.

In the Conclusion part, heuristic remarks about the concept of the « constitutive ether » are made. According to this concept, the quantum aspect of microphysics comes from the fact that the elementary objects are special kinds of flows in an imagined perfect fluid (or « micro-ether »), such as a vortex. On the other hand, gravitation would be a continuous variation, due to the accumulation of a great number of vortices, of the macroscopic pressure in the assumed substratum. This leads the author to *conjecture* that *gravitation is not a quantized interaction*, for it has the nature of a macroscopic correction: even though it affects matter down to the scale of elementary particles, the definition of the gravitational force is macroscopic. The source of the gravitational field is the *classical* energy-momentum tensor (or its energy component).

## 2. Dispersion relations and the quantum correspondence

*2.1 Dispersion equation and wave equation in a (classical) linear medium*

As everyone knows, a general wave $\psi$ may be defined as the product of a « smoothly varying » amplitude $A(t, \mathbf{x})$ and an oscillating part $\phi(\theta(t, \mathbf{x}))$, with $\phi$ a periodic scalar function of the dimensionless *real* variable $\theta$, the latter or « phase » being itself a function of the time $t$ and the position **x**. The « position » **x** belongs to an $N$–dimensional configuration space M, which may be the physical space ($N = 3$). In physics, $\phi$ is usually the complex exponential: $\phi(\theta) = \exp(i\theta)$ (or its real part, $\cos\theta$), but other « wave profiles » do also occur. The notion of a smoothly varying amplitude means that the wave structure may be recognized: to the very least, the relative amplitude variation $\delta A/A$ should not exceed the order of $\delta\theta/T$, with $T$ the period of the wave profile $\phi$. Now if we have a physical law for a given wave phenomenon, it will lead to a partial differential equation for the wave function $\psi$. Let us assume that $\psi$ is a *scalar* and that this equation is *linear:*

$$P\psi \equiv a_0(X)\,\psi + a_1^\mu(X)\,D_\mu\,\psi + \ldots + \sum_{\mu_0+\ldots+\mu_N=n} a_n^{\mu_0\ldots\mu_N}(X)\,(D_0)^{\mu_0}\ldots(D_N)^{\mu_N}\,\psi = 0,$$
(2.1)

where $X$, with coordinates $x^\mu$ ($0 \le \mu \le N$), is the relevant point of the product « time » × configuration space (extended configuration space **R**×M), and $D_\mu\psi = \psi_{,\mu}$ is the partial derivative. Quite often, the order of the equation is $n = 2$:

$$P\psi \equiv a_0(X)\,\psi + a_1^\mu(X)\,\psi_{,m} + a_2^{\mu\nu}(X)\,\psi_{,\mu,\nu} = 0.$$
(2.2)

Given the phase function $\theta(X)$ of a general wave, one defines the wave covector $\mathbf{K} = \nabla\theta = (-\omega, \mathbf{k})$, *i.e.* $K_\mu = \theta_{,\mu}$. In the linear case, it is appropriate to consider a sinusoidal wave profile: $\phi(\theta) = \exp(i\theta)$, hence $\phi_{,\mu} = iK_\mu\phi$. Let us consider a wave function $\psi = A\,\exp(i\theta)$, with $A$ constant, which is an « elementary wave » *at the point X considered*, *i.e.*, such that $K_{\mu,\nu} = 0$ at point $X$. The *necessary and sufficient* condition for such wave function to obey Eq. (2.2) at point $X$ is:

$$\Pi(X,\mathbf{K}) = 0, \qquad (2.3)$$
$$\Pi(X,\mathbf{K}) \equiv \Pi_X(\mathbf{K}) \equiv a_0(X) + ia_1^\mu(X)K_\mu - a_2^{\mu\nu}(X)K_\mu K_\nu,$$

the obvious generalization to Eq. (2.1) being (under the condition that all derivatives of the $K_\mu$'s, up to the order $n-1$, vanish at the point $X$ considered):

$$\Pi(X,\mathbf{K}) \equiv \Pi_X(\mathbf{K}) \equiv a_0(X) + ia_1^\mu(X)\,K_\mu + \ldots$$
$$+ i^n \sum_{\mu_0+\ldots+\mu_N=n} a_n^{\mu_0\ldots\mu_N}(X)\,K_0^{\mu_0}\ldots K_N^{\mu_N} = 0. \quad (2.4)$$

We shall name Eq. (2.3) (or (2.4)) the *dispersion equation* of Eq. (2.2) (or (2.1)). If one makes any coordinate change $x'^\rho = x'^\rho(x^\mu)$ in Eq. (2.2), the coefficient $a_0(X)$ is left unchanged and the $a_1^\mu(X)$ series transforms like a vector. However, the $a_2^{\mu\nu}(X)$ series transforms like a contravariant second-order tensor, if and only if one makes an « *infinitesimally linear* » coordinate change, *i.e.*, if $\partial^2 x'^\rho/\partial x^\mu \partial x^\nu = 0$ at the point considered. Hence, the left-hand side of the dispersion equation (2.3) is an invariant scalar only under infinitesimally linear coordinate transformations. Similarly, the condition $K_{\mu,\nu}(X) = 0$, used to derive (2.3) from (2.2), is covariant only under infinitesimally linear coordinate transformations. This means that, at each point $X$ of the extended configuration space **R**×M, a linear group of privileged coordinates must be available among the local coordinates valid in a neighborhood of $X$. Hence, **R**×M must be equipped with a more particular structure than just that of a differentiable manifold. At this point, the special role of the time coordinate does not appear compelling. A pseudo-Riemannian metric $\gamma$ on **R**×M is hence enough: we then select the group of the locally geodesic coordinate systems (LGCS) at $X$ for $\gamma$ [$\gamma_{\mu\nu,\rho}(X) = 0$ for all $\mu$, $\nu$ and $\rho$], two of which exchange indeed by an infinitesimally linear transformation. Thus, in order to evaluate the dispersion equation (2.3) at $X$, we can take any LGCS and we get the polynomial function $\Pi_X$ of covector **K**; this function does not depend on the LGCS. Hence, the function $\Pi$ (of $X$ and **K**, thus a function defined on the « cotangent bundle », T*(**R**×M), to **R**×M) is well-defined. In the sequel we shall need the « projection time » $t$ as a preferred time coordinate (up to a constant factor: $x^0 = \alpha t$).

Similarly, for an equation of order $n \geq 3$, the dispersion equation (2.4) is an invariant only under coordinate changes whose all derivatives, up to the order $n$, are zero at the point $X$ considered. The condition that all derivatives of the $K_\mu$'s, up to the order $n-1$, vanish at $X$, is also covariant under those changes only. And if we have a pseudo-Riemannian metric $\gamma$, we might hope to define a privileged class of coordinates as the « $n$-LGCS systems », *i.e.* those in which all derivatives of the metric, up to the order $n-1$, are zero: this condition is stable under the changes just mentioned. However, already for $n = 3$, a « 3-LGCS system » exists only if the Riemann tensor vanishes at $X$. Hence, *except for part 3, which does not depend on the correspondence discussed in part 2, the results presented in this paper are valid only for equations of order $n \leq 2$, or also for the case of a flat space-time metric.* This restriction has few practical consequences, obviously.

If the wave equation (2.1) has constant coefficients (a notion that is covariant only under truly linear coordinate changes, hence implies a vector space structure), it may indeed have solutions that are an « elementary wave » at *each point X* in an open domain of $\mathbf{R} \times M$ (such solutions have a constant wave covector, hence are plane waves). But the dispersion equation makes sense for a general linear wave equation. In order that the wave covector be *real*, one demands that the coefficients of the dispersion equation are real (it is at this point that the general linear equation (2.1) really becomes a wave equation). This means that, for a *real* wave equation, only odd derivatives, *or* only even derivatives, are allowed [21]. However, one may mix odd and even derivatives, provided one accepts complex coefficients in the wave equation – as for Schrödinger's equation.

Now we state the essential result of this Section: the correspondence between the linear operator P and the function $\Pi$ is one-to-one, in other words one may *uniquely* pass from the wave equation to the dispersion equation *and conversely*. This is undoubtedly true, since the data of either P or $\Pi$ is equivalent to the data of the set of the scalar functions $a_0$, $a_1^\mu$, $a_2^{\mu\nu}$, and so on, of the point $X$ in $\mathbf{R} \times M$. The important point is that this result is true also in the general case with *variable* coefficients, in which $a_0$, $a_1^\mu$, $a_2^{\mu\nu}$, ..., do depend on $X$. In the case with constant coefficients, the inverse correspondence, from (2.4) to (2.1), amounts to the substitution

$$K_\mu \to \frac{1}{i} D_\mu. \qquad (2.5)$$

This remains true in the case with variable coefficients, provided one orders each monomial in the dispersion equation as in (2.3) or (2.4), *i.e.*, « $X$ before $\mathbf{K}$ ». This ordering is the natural ordering for a polynomial in $\mathbf{K}$ with coefficients that are functions of $X$. It corresponds to the ordering in Eq. (2.1) or (2.2), which also is the natural ordering for the *general* linear differential equation: the (variable) coefficient comes before the differentiation.

*2.2 Group velocity and Hamiltonian motion*

As mentioned above, we now need the preferred « projection » time $t$, with $X = (t, \mathbf{x})$, and we seek to compute the frequency $\omega = -K_0$ as a function of the « spatial » part $\mathbf{k}$ of $\mathbf{K}$: $\omega = \omega(k_1, ..., k_N) = \omega(\mathbf{k})$. When extracting the real roots of the polynomial equation $\Pi_X(\mathbf{K}) = 0$, considered as an equation for the unknown $K_0$ with the data $k_1, ..., k_N$, one may follow the different roots $W_1, ..., W_n$ (at most $n$) as functions of $X$. It is assumed in the following that one such particular root has been identified, and we have thus:

$$\omega = W(k_1, ..., k_N; X) = W(\mathbf{k}; X) = W(\mathbf{k}, \mathbf{x}, t), \quad (2.6)$$

which is called « the » *dispersion relation*, it being kept in mind that several different such relations may in general be extracted from the unique dispersion *equation* [21]. Of course, $W$, which will be called the *dispersion*, is in general not a polynomial function of $\mathbf{k}$ at fixed $X$. Now let us assume that some wave function of the general form

$$\psi(t, \mathbf{x}) = A(t, \mathbf{x}) \exp(i\,\theta(t, \mathbf{x})), \quad (2.7)$$

is such that the corresponding wave covector obeys exactly the dispersion relation (2.6) (we do *not* assume here that $\psi$ is an exact solution of the wave equation (2.1), and indeed this is in general incompatible with our assumption). One defines the *group velocity* for this wave function as the « spatial » vector with components

$$C^j = C^j(\mathbf{k}, \mathbf{x}, t) = \frac{\partial W}{\partial k_j}(\mathbf{k}, \mathbf{x}, t) \quad (2.8)$$

(Latin indices will be reserved for « spatial » components). This is the natural generalization [21] of the usual notion of group velocity in a uniform medium. From the definitions $K_\mu = \theta_{,\mu}$ and $\omega = -K_0$, and from the symmetry of the second derivatives of $\theta$, one gets

$$\frac{\partial \omega}{\partial x^j} + \frac{\partial k_j}{\partial t} = 0, \quad \frac{\partial k_i}{\partial x^j} = \frac{\partial k_j}{\partial x^i}, \quad (2.9)$$

and since here $\omega(t, \mathbf{x}) = W(\mathbf{k}(t, \mathbf{x}), \mathbf{x}, t)$ by hypothese, Eq. $(2.9)_1$ may be rewritten as [21]

$$\frac{\partial W}{\partial x^j} + C^i \frac{\partial k_j}{\partial x^i} + \frac{\partial k_j}{\partial t} = 0. \quad (2.10)$$

This is a hyperbolic equation for the (spatial) wave covector $\mathbf{k}$, showing that $\mathbf{k}$ propagates with the group velocity. It is immediate to verify that it can be put in the following characteristic form [21]:

$$\frac{dk_j}{dt} = -\frac{\partial W}{\partial x^j} \text{ on curve } \frac{dx^j}{dt} = C^j = \frac{\partial W}{\partial k_j}. (2.11)$$

Thus, the motion of the wave vector in a general nonuniform medium is governed by a Hamiltonian dynamical system, the Hamiltonian being the dispersion $W$. To the author's limited knowledge of the literature on quantum mechanics, this crucial result has been overlooked there.

*2.3 The quantum correspondence and its (non-)ambiguity*

Let us consider a classical system of point particles with « position » $\mathbf{x}$ (in the configuration space M), its dynamics being given by a Hamiltonian system with Hamiltonian $H$ and conjugate momentum $\mathbf{p}$ (a « spatial » covector):

$$\frac{dp_j}{dt} = -\frac{\partial H}{\partial x^j}, \quad \frac{dx^j}{dt} = \frac{\partial H}{\partial p_j}. \quad (2.12)$$

Following de Broglie and Schrödinger, let us imagine that those microscopic objects, which we initially described as point particles, actually have a spatial structure made of some (unknown) waves; and that this classical Hamiltonian dynamics only describes the « skeleton » of the wave motion, in *precisely the same way* as geometrical optics describes the trajectories of « light rays », which constitute the skeleton of the light wave propagation pattern. Thus, we expect that the dynamics (2.12) should be that approximation of the wave motion which becomes exact at the « nil wave length » limit where, in the neighborhood of any point $X$ in the extended configuration space, the wave may be considered as a plane wave, that is

$$A \approx \text{Const. and } \delta\theta \approx \mathbf{k}.\delta\mathbf{x} - \omega\delta t = K_\mu\, \delta x^\mu \quad (2.13)$$

in Eq. (2.7); of course, condition $(2.13)_2$ means that the higher-order derivatives of $\theta$ can be neglected: $K_{\mu,\nu} \approx 0$, and so on. We seek for the linear wave equation (2.1) for which the Hamiltonian dynamics would represent just that approximation. Assume that the wave function $\psi$, of the form (2.7), is the relevant *solution*, in a situation where the Hamiltonian dynamics is a good approximation, to that *exact* wave equation: in view of what we have just said, $\psi$ will satisfy the approximate conditions (2.13). But the exact conditions $A = \text{Const}$, $K_{\mu,\nu} = 0$, and so on, are the conditions under which the substitution of (2.7) into (2.1) gives the dispersion equation (2.4). Hence, the wave covector $\mathbf{K}$ of an exact solution (2.7) to the wave equation (2.1) satisfies the dispersion equation (2.4) to an approximation which becomes exact in the « geometrical optics » limit where the dynamics (2.12) itself becomes exact.

Now, a *regular* solution of the dispersion equation (2.4) should correspond to a unique « branch », *i.e.*, it should satisfy one and only one among the different possible dispersion relations (2.6), and we assume that it is indeed the case. Therefore, in the « geometrical optics » limit, the spatial wave covector **k** [associated with a wave function (2.7) obeying the wave equation (2.1)] obeys also the « Hamiltonian dynamics » (2.11): the only difference with a true Hamiltonian dynamics is that, in Eq. (2.11), a *continuous distribution* of Hamiltonian trajectories is involved. However, to account for the fact that (in its domain of relevance), the classical approximation (2.12) successfully considers point particles, we expect that, at least in some cases, the wave pattern is made of spatially concentrated « wave packets », in the exterior of which the amplitude is negligibly small. Obviously, *we are led to admit that the Hamiltonian systems characterized respectively by the dispersion W (Eq. (2.11)) and by the classical Hamiltonian H (Eq. (2.12)) have exactly the same set of possible trajectories* $\mathbf{x} = \mathbf{x}(t)$.

This can be achieved by a *canonical* transformation $(\mathbf{x}, \mathbf{k}, t, W) \to (\mathbf{x}, \mathbf{p}, t, H)$ (thus with the « position » and time variables being left unchanged) only if $\mathbf{k} = \mathbf{p}$ and $W = H$. But a canonical transformation means that the trajectories in the extended phase space, $\mathbf{R} \times T^*M$, are the same: here we merely want that their *projections* in the extended configuration space $\mathbf{R} \times M$ are the same. If we consider the Lagrangians $\Lambda(\mathbf{x}, \dot{\mathbf{x}}, t)$ and $L(\mathbf{x}, \dot{\mathbf{x}}, t)$ respectively associated with $W$ and $H$ by the Legendre transformation, the condition is that the extremals of the respective action integrals must be the same. The obvious, simplest way to ensure that this is true, is to impose that $\Lambda$ and $L$ are equal up to a multiplicative constant. We denote this constant by $\hbar$... Thus, we are led to admit that $L = \hbar \Lambda$, whence follows that

$$H = \hbar W, \text{ or } E = \hbar \omega, \qquad (2.14)$$

and for the canonical conjugate momenta:

$$\frac{\partial L}{\partial \mathbf{x}} = \hbar \frac{\partial \Lambda}{\partial \mathbf{x}}, \quad \text{or} \quad \mathbf{p} = \hbar \mathbf{k}. \qquad (2.15)$$

The *a priori* interpretation of the correspondence (2.14) and (2.15) is a *formal* one: it is the correspondence between the Hamiltonian systems (2.11) and (2.12) – that govern respectively (i) the bicharacteristics of the propagation equation (2.10) for a wave vector obeying the dispersion relation (2.6), and (ii) the trajectories of a classical system– under the condition that the « geometrical optics » limit of the wave equation (2.1) gives the same trajectories as the classical system. Moreover, let us recall that the wave equation can be *uniquely* deduced from the dispersion *equation* (2.4) (see the end of §2.1). However, the correspondence (2.14) gives us the dispersion *relation* (2.6). But the dispersion relation is one root of the polynomial dispersion equation (2.4), the latter being seen as an equation for $\omega = -K_0$.

Thus, if we are given a classical Hamiltonian $H$ and if we seek for the associated « quantum » wave equation, we first (trivially) deduce the dispersion $W$ by the correspondence (2.14) and (2.15), *i.e.*

$$W(\mathbf{k}, \mathbf{x}, t) = [H(\hbar \mathbf{k}, \mathbf{x}, t)]/\hbar. \qquad (2.16)$$

Then: *either* (i) $H$ is a polynomial in **p** (at fixed $X$), such as, for instance,

$$H(\mathbf{p}, \mathbf{x}, t) = \frac{\mathbf{p}^2}{2m} + V(\mathbf{x}, t). \qquad (2.17)$$

In this case, the data of $H$ (or $W$) is simply the dispersion equation: more exactly, it is the simplest (lowest degree) dispersion equation that has $W$ as one possible dispersion. Therefore, the correspondence (2.5) gives unambiguously the wave equation. This is also true if the coefficients of the polynomial *depend* on **x** and *t*. Note that the composition of the two correspondences (*E*, **p**) → (ω, **k**) [Eqs. (2.14) and (2.15)] and (2.5) gives the « quantum correspondence » :

$$E \to +i\hbar \frac{\partial}{\partial t}, \quad p_j \to -i\hbar \frac{\partial}{\partial x^j} \qquad (2.18)$$

(from Eq. (2.17), one thus gets Schrödinger's equation). *Or* (ii) *H* is not a polynomial in **p**, but some polynomial function of *H* is itself a polynomial in **p**. In that case, we take the dispersion equation as that polynomial function *P* which has the lowest degree, which is necessarily unique up to a constant factor, and we thus obtain uniquely the linear wave equation with the lowest possible order. In practice, the Hamiltonian *H* will then appear precisely as involved in such polynomial function: this is the case for the relativistic Hamiltonian of, for instance, a free particle, which appears in the equation

$$[H(\mathbf{p})]^2 - \mathbf{p}^2 c^2 - m^2 c^4 = 0. \qquad (2.19)$$

For the latter dispersion equation (up to the correspondence (2.14) and (2.15)), the wave equation obtained by the correspondence (2.5) is, of course, the Klein-Gordon equation, with the d'Alembert equation as the case *m* = 0. The latter case shows that the approach based on the « geometrical optics » limit does work in the genuine case! It also recalls to us that *the method gives merely the wave equation for anyone scalar component, and tells us nothing about the scalar or tensor character of the physically relevant wave field.*

To be complete, there is still the case that (iii) *H* is not an algebraic function of **p**, that is, no polynomial function of *H* becomes a polynomial in **p**. This means simply that no linear wave equation can be associated with the given classical Hamiltonian by this correspondence.

*2.4 Comments on the physical meaning*

The usual correspondence (2.18), between a classical Hamiltonian and a « quantum » wave equation, may thus be arrived at by a seemingly new method, based on the dispersion relation in a linear medium. The interest one may find to this method is that it gives a rather rigorous basis to the essence of wave mechanics: the idea that classical dynamics of point particles should correspond to the « geometrical optics » limit of a linear wave equation. It has been shown that this idea leads, modulo the study of classical dispersion relations [21], to the correspondence (2.18) in a natural and general way. The method is not limited to the case of a single particle: it works for any Hamiltonian system. Moreover, the method also indicates how to make the correspondence unambiguous in the case where the classical Hamiltonian contains mixed terms involving both the momentum **p** and the (configuration-)space–time position *X* = (*t*, **x**). The rule to obtain the wave operator unambiguously is simple: just put the function of *X* as a multiplying coefficient before the monomial in **p** – as the dispersion equation has to be a polynomial in **p** at fixed *X*. This is important in the case with a gravitational field (part 4 of this paper), for a « curved space-time » means that, for example, the kinetic energy becomes such mixed term (since it involves the metric tensor, that depends on *X* ).

Thus, a classical Hamiltonian (even in an implicit form such as Eq. (2.19)) gives unambiguously a linear wave equation for the more complex wave structure supposed to underly the approximate classical behaviour of a system of particles (the latter corresponding to the nil wave length limit). However, if one admits that it is a wave structure which is the fundamental behaviour, then one must expect, of course, that in some cases no classical Hamiltonian will be relevant, *i.e.*, one must expect that the simple correspondence (2.18) is not always a sufficient tool to obtain a correct wave equation – an example is Dirac's equation.

If one admits that the quantum wave function $\psi$ describes a physical wave, then one should be able to define a local energy density $\rho$ for the wave, for which a local conservation should apply, involving also the force fields to which the quantum object is subjected (and which, conversely, the quantum object should in general influence). The « space » integral of $\rho$ would be the total energy of the quantum object, which should be conserved if it is in a constant force field, and possibly also under more general conditions. For example, if the force field is electromagnetic, the energy of the field is conserved if the field lines of the Poynting vector field are closed, hence the possibility of « purely electromagnetic » quantum particles (see Vlasov [18] and Waite [19]). Thus, the energy density $\rho$ would be substituted for the probability density of standard QM. Now, for the one-particle Schrödinger equation, there is the well-known local conservation equation, that involves only the wave function $\psi$, and which expresses the conservation of $|\psi|^2$. Hence, one is tempted to assume $\rho = |\psi|^2$ in that case ... However, for other equations such as the Klein-Gordon equation, $|\psi|^2$ is not a conserved quantity, although the energy density of the Klein-Gordon field (the $T^0_0$ component of the associated energy-momentum tensor **T**) is well-defined, positive and conserved [6]. From the tentative point of view of a « mechanics of real microphysical waves », there does not seem to be any compelling reason to admit in general that the wave function $\psi$ should be square-integrable and that the integral of $|\psi|^2$ should be a conserved quantity – although there is indeed such reason for the usual one-particle Schrödinger equation. However, as long as the wave function is supposed to represent a single quantum particle (which is in principle impossible in the relativistic case, since the particle number is not conserved then), its spatial extension should be limited in some sense: for example, a function $\psi$ such that $|\psi|$ is constant in space should not be a correct wave function for a single particle.

## 3. Eigenvalues problems for energy states and for momentum states

The quantum-mechanical aspect in the correspondence between a Hamiltonian and a wave equation lies in the wave-particle duality: thus, a system of mass particles should be handled in general as a wave function defined on the extended configuration space and, conversely, a classical electromagnetic field obeying the d'Alembert equation should be treated, in the geometrical optics limit, as consisting of zero-rest-mass particles. However, the genuinely « quantum » (*i.e.*, discrete) aspect of quantum mechanics is the role played by discrete eigenvalue spectra. It seems desirable that, as well as the above correspondence, also the eigenvalue problems should occur in a natural way: the best would be to avoid the *a priori* postulate that « the quantum observables are operators in a Hilbert space and the measurement operations lead to eigenvalues for these operators ». In this question, one should not limit oneself to wave equations that can be obtained by the correspondence discussed above.

### 3.1 Energy states and stationary waves

As it is well-known, the quantization of the energy levels appears naturally, in Schrödinger's wave mechanics, when one looks for stationary solutions of his equation. More generally, let H be a « quantum Hamiltonian », *i.e.*, a linear operator acting on functions $\psi$ defined on the extended configuration space $\mathbf{R} \times \mathbf{M}$, and assume that it does not contain the time.[1] For a such « purely spatial » quantum Hamiltonian, one searches for stationary

---

[1] In precise terms: H involves only « spatial » derivatives (*e.g.* in Eq. (2.2) with H in the place of P: $a_1^0(X) = 0$, and $a_2^{\mu\nu}(X) = 0$ if $\mu = 0$ or $\nu = 0$), *and* its coefficients $a_0(X)$, $a_1^\mu(X)$, $a_2^{\mu\nu}(X)$, ..., actually depend merely on **x**, not on $t$. (This will be the case, *e.g.*, if H is deduced by the quantum correspondence $(2.18)_2$ from a classical Hamiltonian $H(\mathbf{p}, \mathbf{x})$, polynomial in **p** and time-independent.)

solutions of the following Schrödinger-type equation:[2]

$$\mathrm{H}\psi = i\hbar\frac{\partial \psi}{\partial t}. \qquad (3.1)$$

A precise definition of stationarity may be that the time and « space » variables are separated:

$$\psi(t, \mathbf{x}) = \phi(t)a(\mathbf{x}), \qquad (3.2)$$

and that the time-dependence $\phi$ is a quasi-periodic function. Putting (3.2) in (3.1) gives

$$\phi(t)\mathrm{H}a(\mathbf{x}) = i\hbar\phi'(t)a(\mathbf{x}), \qquad (3.3)$$

hence the only possibility is that $\phi'/\phi = \mathrm{Const} \equiv -i\omega$, whence

$$\phi(t) = A\exp(-i\omega t), \quad \mathrm{H}a = \hbar\omega a \equiv Ea, \qquad (3.4)$$

with $\omega$ real in order that $\phi$ be quasi-periodic (and indeed periodic). Thus, the spatial part $a$ is a solution of the eigenvalue problem $(3.4)_2$, while the time part is sinusoidal with frequency $\omega = E/\hbar$, where $E$ is the eigenvalue. Note that the elementary proof of this *well-known result* remains true in the case where the wave function $\psi$ takes values that are not scalars, but elements of some *vector space* (*e.g.* a spinor space), provided one demands that the time-dependence, $\phi(t)$ in Eq. (3.2), is a scalar. Hence, it applies to Dirac's equation.

Thus, as long as the wave equation is of the general Schrödinger type (3.1), one may dispense oneself from the *a priori* assumption that the energy levels are eigenvalues of the Hamiltonian operator, and replace it by the assumption that the energy levels correspond to stationary states. But what does an energy level mean? What is experimentally observed is a frequency spectrum ($\omega'_n$) of electromagnetic radiation, so one should postulate, more strictly, that any observed frequency $\omega'$ of the emitted/absorbed radiation is the absolute difference between two eigenfrequencies:

$$\omega' = |\omega_1 - \omega_2| = |E_1 - E_2|/\hbar. \qquad (3.5)$$

This is the Bohr relation, which was proposed in the framework of the light quanta hypothese, but it seems to be interpretable with a « real wave » picture, as Schrödinger suggested [14-15]. It then seems to indicate that there is some identity of nature between the « matter waves » and the electromagnetic waves, since the frequencies suffer no distortion. This would be understandable if all waves were in a common substratum or ether. However, the well-known difficulty for such a realistic interpretation is that the matter waves are in a configuration space.

In deriving the conclusion (3.4) that a stationary wave has a sinusoidal time-dependence and that its amplitude $a(\mathbf{x})$ is the solution of an eigenvalue problem, no restriction is imposed on the spatial part of the wave equation, but the time part is assumed to be the r.h.s. of Eq. (3.1), thus of the first order. It is interesting to see what happens with the (free) Klein-Gordon (K-G) equation, which is second-order in time:[3]

$$\Delta\psi - \frac{1}{c^2}\frac{\partial^2\psi}{\partial t^2} - \frac{1}{\lambda^2}\psi = 0, \quad \lambda \equiv \frac{\hbar}{mc}. \qquad (3.6)$$

With the ansatz (3.2), it becomes

---

[2] A such equation will be obtained, *e.g.*, from the classical $H(\mathbf{p}, \mathbf{x})$, by using now the correspondence $(2.18)_1$.

[3] In textbooks, it is often objected to the K-G equation that a quantum wave equation should be first-order in time. But since wave functions with *various* numbers of scalar components occur in QM (*e.g.* Schrödinger scalars *vs.* Dirac 2-spinors), this restriction does not have an objective meaning. In the derivation of Dirac's equation, the restriction imposed by relativity is that the order of derivation should be the same for the space variables as for the time variable. The choice of writing a first-order equation (acting, in the case of Dirac's equation, on wave functions with an *a priori unknown* number of scalar components), is in general a matter of convenience. As a matter of fact, the K-G equation may be rewritten in the Schrödinger form (3.1), by introducing a 2-component wave function [9]. However, it is more expedient here to stay with the scalar form (3.6).

$$\phi \, \mathrm{U}a = (\phi\,''/c^2)\, a, \qquad \mathrm{U}a \equiv a - a/\lambda^2, \quad (3.7)$$

hence $\phi\,''/\phi = \text{Const} \equiv -\varepsilon\omega^2$, $\varepsilon = \pm 1$. The only quasi-periodic solution $\phi$ is periodic ($\varepsilon = 1$):

$$\phi = A \exp(\mathrm{i}\omega t) + B \exp(-\mathrm{i}\omega t), \quad (3.8)$$

with $\omega$ real, and $a$ is a solution of the eigenvalue problem

$$-\Delta a = \left(\frac{\omega^2}{c^2} - \frac{1}{\lambda^2}\right) a. \quad (3.9)$$

For « confined » boundary conditions, the problem $-\Delta a = \mu a$ admits a discrete sequence $(\mu_n)$ of positive eigenvalues. With each of them, one may associate one « Schrödinger frequency », $\omega_n = \mu_n \hbar / 2m$, and two « K-G frequencies », $\omega'_n{}^+ = c(\mu_n + m^2 c^2/\hbar^2)^{1/2}$ and $\omega'_n{}^- = -\omega'_n{}^+$. One finds that the difference between two K-G frequencies, $\omega'_n{}^+ - \omega'_l{}^+$, is approximately equal to $\omega_n - \omega_l$, provided the « Schrödinger energies », $E_n = \hbar\omega_n$ and $E_l = \hbar\omega_l$, are negligible as compared with the rest-mass energy $mc^2$. So the non-relativistic limit works well.

*3.2 Momentum states: a generalization of plane waves*

In QM, it is admitted that a (single) quantum particle has a well-defined momentum **p**, if and only if its wave function (in the physical space, which coincides, in the case of a single particle, with the configuration space) is that of a plane wave. Moreover, only the spatial dependence of the wave function is envisaged, so that the « wave function » (at a given time) is $A \exp(\mathrm{i}\, \mathbf{k}\cdot\mathbf{x})$. Thus, the first wave function encountered in QM has constant modulus and, in particular, it is not square-integrable, although for the case of one quantum « particle » obeying Schrödinger's equation, this is a necessary requirement (*cf.* the end of § 2.4). Actually, it is easy to show that a plane wave cannot obey the time-dependent Schrödinger equation unless the potential $V$ is a constant number (thus unless the « particle » is in a free motion); this will be proved in passing below. Perhaps one may ask if the wave function of a *moving* particle could be defined as a function of *time* also, and if its spatial extension could be limited in some sense.

Thus, we consider a wave function of the general form (2.7), and we ask under which particular conditions it could be taken to represent a quantum object having, at any given time $t$, a well-defined momentum $\mathbf{p}(t)$. In agreement with the notion that the velocity of the particle is related to the group velocity of the associated wave (Eq. (2.11)), we impose a first requirement: (i) *the group velocity* **C** *must be a function of t only*. This seems to be a natural condition. However, **C** depends on the dispersion $W$ (Eq. (2.8)). The latter is found by the correspondence (2.14), in the particular case that the quantum wave equation can be deduced from a classical Hamiltonian by the correspondence (2.18). In the general case, the dispersion equation is found directly from the wave equation by the transition from Eq. (2.1) to Eq. (2.4), thus by the substitution $D_\mu \to \mathrm{i}\, K_\mu$, provided the wave equation can be put in a form where it is the same for each component of the wave function; then, several dispersions (whence several **C** vectors) may be defined by extraction (see § (2.2)). In any case, the definition of **C** depends on the wave equation. Hence, the natural condition (i) implies that *the very definition of a « pure momentum state » should depend on the quantum wave equation*. Let us assume that this is the one-particle Schrödinger equation, associated with the non-relativistic Hamiltonian (2.17). The group velocity is then

$$\mathbf{C} = \hbar \mathbf{k}/m, \quad (3.10)$$

so that, in the case of the usual Schrödinger equation, condition (i) is equivalent to the requirement that (i)' *the spatial wave covector* **k** *is a function of t only* [4]. Thus $\nabla_\mathbf{x}\theta = \mathbf{k}(t)$, which is equivalent to say that the phase $\theta$ has the form

---

[4] This is not true in general, *e.g.* it is wrong for the (scalar) Schrödinger equation in the presence of a magnetic vector

$$\theta(t, \mathbf{x}) = \mathbf{k}(t).\mathbf{x} - f(t). \quad (3.11)$$

Note that, under condition (i)', the frequency is not necessarily uniform:

$$\omega = f'(t) - \mathbf{k'}(t).\mathbf{x}. \quad (3.12)$$

Obviously, condition (3.11) is fulfilled by a plenty of possible wave functions, since no restriction is imposed on the amplitude *A*. But since we recognized that our definition of a momentum state depends on the wave equation, we have one more reason to impose on the wave function $\psi$ the condition that (ii) *$\psi$ should obey the wave equation* – which is anyway a necessary requirement [5]. Furthermore, we assume from now on that the amplitude *A*, as well as the phase $\theta$, is *real*. This is the necessary condition under which the writing of the complex function $\psi$ in the form (2.7) is (practically) unambiguous: otherwise, the relation between $\psi$ and its phase $\theta$ would be *totally* ambiguous. (The restriction that *A* also should be real does not play a role in the study of dispersion relations, since one then considers constant amplitudes.) After an easy algebra, the time-dependent Schrödinger equation for the wave function (2.7), with the phase (3.11), is then found equivalent to the two real equations:

---

potential **A**: in that case, one finds $\mathbf{C} = [\hbar\mathbf{k}-(q/c)\mathbf{A}]/m$ (consistently with the fact that, for a classical particle governed by the relevant Hamiltonian, $\mathbf{P} - (q/c)\mathbf{A} = m\dot{\mathbf{x}}$ is the « true » momentum, where **P** is the canonically conjugate momentum, to which applies the relation $\mathbf{P} = \hbar\mathbf{k}$). Hence, $\mathbf{k} = \mathbf{k}(t)$ is then *incompatible* with $\mathbf{C} = \mathbf{C}(t)$ unless $\mathbf{A} = \mathbf{A}(t)$ (*i.e.,* no magnetic field).

[5] In the case where a magnetic field is present, the expression of the group velocity involves **A** that depends on the gauge condition (note 4). Therefore, one is really enforced to impose the wave equation: if one changes the gauge condition, the wave function, hence also the wave covector **k**, will then transform according to a simple rule [11], and it is easy to verify that the group velocity is then independent of the gauge.

$$\mathrm{H}A \equiv -\frac{\hbar^2}{2m}\Delta A + VA = \left(\hbar\omega - \frac{\mathbf{p}^2}{2m}\right)A, \quad (3.13)$$

$$\mathbf{p}(t) \equiv \hbar\mathbf{k}(t),$$

$$\frac{dA}{dt} \equiv \frac{\partial A}{\partial t} + \mathbf{C}.\nabla A = 0. \quad (3.14)$$

Hence, we may wish to impose the additional requirement that (iii) *the dispersion relation is satisfied*:

$$\hbar\omega - \frac{\mathbf{p}^2}{2m} - V = 0, \quad (3.15)$$

and this requirement implies that Eq. (3.13) is equivalent to the Laplace equation:

$$\Delta A = 0. \quad (3.16)$$

However, Eq. (3.15), in which we recall that $\mathbf{p} = \mathbf{p}(t) \equiv \hbar\mathbf{k}(t)$, implies also that *V* is a linear function of **x**. Indeed, since $\omega$ is given by Eq. (3.12), we deduce from (3.15) that

$$V(\mathbf{x}, t) = \hbar f'(t) - \hbar\mathbf{k'}(t).\mathbf{x} - \frac{\hbar^2}{2m}\mathbf{k}^2. \quad (3.17)$$

Conversely, if we know that $\Delta A = 0$, then, unless $A(t, \mathbf{x}) = 0$, Eq. (3.13) is equivalent to Eq. (3.15) and thus implies Eq. (3.17). In particular, a plane wave ($A$ = Const, $\mathbf{k}$ = Const, $\omega$ = Const) can be a solution of the Schrödinger equation only if the potential *V* is a constant, as announced. This result, plus the fact that a plane wave has a constant modulus over the space, seems to imply that *a plane wave cannot represent a correct description of the objective state of a single quantum object having a well-defined momentum*. On the other hand, the more general result expressed by Eq. (3.17) is that the dispersion relation cannot be verified for a momentum state which is an exact solution of the Schrödinger equation, unless the potential *V* is linear in space, *i.e.*, unless the force field **F** is uniform. What is more: if *V* is not linear so that we cannot exactly impose the dispersion relation, it seems that the time evolution of the wave covector, hence of the « momentum » $\mathbf{p}(t) \equiv \hbar \mathbf{k}(t)$, is entirely arbitrary – and this appears to be a serious difficulty. Whereas, if the potential has the form (3.17), the evolution of $\mathbf{p}$ is simple and natural:

$$\frac{d\mathbf{p}}{dt} \equiv \hbar \frac{d\mathbf{k}}{dt} = -\nabla V \equiv \mathbf{F}(t). \quad (3.18)$$

Therefore, one is inclined to interpret the foregoing result thus: a « pure momentum state » should satisfy conditions (i), (ii) *and* (iii). *I.e.*, the group velocity should be uniform, and the wave equation *plus* the dispersion relation should be satisfied. One then finds that, for the usual Schrödinger equation, what is called here *a pure momentum state can exist only over space and time intervals such that the traversed force field may be considered as spatially uniform.*

Thus, considering a uniform force field $\mathbf{F}(t)$, let us study such a pure momentum state precisely. Given the initial momentum $\mathbf{p}_0$, its time evolution is completely determined by *Newton's second law* (3.18). Due to Eq. (3.10) with $\mathbf{k} = \mathbf{k}(t)$, the momentum may really be considered as uniformly distributed in the wave. The dispersion relation (3.15), *i.e.*, « $E = \hbar\omega$ », determines the frequency $\omega$. Note that it depends on space and time. The group velocity being known, Eq. (3.14) shows that the amplitude $A$ undergoes a translation with the uniform velocity $\mathbf{C}(t)$, *i.e.*, $A$ simply follows the spatially uniform motion. In particular, $A$ is determined at any time if it is known initially. The spatial variation of $A$ is determined by the Laplace equation (3.16). Here, we wish that (a) $A$ is defined *in the whole space* (except perhaps at a *point* singularity) and harmonic, and (b) $A$ tends towards zero as $r \equiv |\mathbf{x}| \to \infty$ (this is a minimum requirement of spatial limitation). First, these conditions imply that there is a ball $|\mathbf{x}| \leq r_1$ such that $A$ is not bounded in this ball, since otherwise it would be harmonic except perhaps at one point, and bounded in the whole space, hence [7, p. 305] it would be harmonic in the whole space; being harmonic and bounded in the whole space, it would be a constant, hence zero by condition (b). Hence, $A$ is not harmonic at one point of the ball $|\mathbf{x}| \leq r_1$ (since otherwise it would be continuous, hence bounded, in this ball), which we choose as the origin. Thus, *A has a point singularity* at $\mathbf{x} = 0$. Now if we impose the subsidiary condition that $A$ is not more singular than $1/r$, *i.e.*, $rA$ is bounded as $r \to 0$, then we get by the same theorem [7, p. 305] that there is a constant $R$ such that $A - R/r$ is a harmonic function $\phi$ on the whole space. But $\phi$ is bounded on the whole space since it is bounded in every bounded domain and since, as $A$ and $1/r$, it tends towards zero as $r \to \infty$. Hence, $\phi$ is a constant $a$, which must be zero. Thus, *conditions* (a) *and* (b) *imply that $A = R/r$* with $R$ an arbitrary constant and $r \equiv |\mathbf{x}|$ after the appropriate choice of the origin $\mathbf{x} = 0$, *unless* one would accept that $A$ be more singular than $1/r$. Condition (b) gives $A \approx R/r$ as $r \to \infty$, *independently* of the latter condition [7, p. 314]. Unfortunately, $R/r$ is not square-integrable. Note that, if we replace condition (b) by the less severe condition that (b)' *A is bounded as $r \to \infty$*, just the same arguments lead to $A = \alpha + R/r$. Now let us consider, for $R$ and $a > 0$, the function

$$A_{R\alpha}(\mathbf{x}) = -\alpha + R/r \quad \text{if } 0 < r \leq R/\alpha,$$
$$A_{R\alpha}(\mathbf{x}) = 0 \quad \text{if } r \geq R/\alpha. \quad (3.19)$$

From a physical point of view, this continuous function is undistinguishable from $R/r$ if $\alpha$ is small enough, and it has a finite square integral. Of course, $A$ does not satisfy $\Delta A = 0$ on the sphere $r = R/\alpha$, but $A$ vanishes from that sphere and, in traversing the sphere, its derivatives undergo an arbitrarily small discontinuity if $\alpha$ is small enough. We note finally that different amplitudes functions $A$ may be found if one accepts that $A$ has, for instance, a one-dimensional singularity (such as a circle, as for a torus vortex). However, as long as one demands that $A$ vanish at infinity, a singularity is unavoidable since $A$ must be a harmonic function.[6]

*3.3 Comments on the physical interpretation of the « momentum states »*

Thus, *in a uniform force field* $-\nabla V = \mathbf{F}(t)$, *the Schrödinger equation admits the wave functions*

$$\psi(t, \mathbf{x}) = A_{R\alpha}[\mathbf{x} - \mathbf{a}(t)] \exp\{i[\mathbf{k}(t)\cdot\mathbf{x} - f(t)]\},$$
$$d\mathbf{a}/dt = \hbar\mathbf{k}(t)/m, \qquad (3.20)$$

*as square-integrable pure momentum states* (see the proposed definition after Eq.(3.18)). The wave covector $\mathbf{k}(t)$ and the function $f(t)$ are determined (up to their arbitrary initial value) by the potential $V(\mathbf{x}, t)$; in particular, the momentum $\mathbf{p}(t) = \hbar\mathbf{k}(t)$ evolves according to Newton's second law (3.18). The *arbitrary* numbers $R$ and $\alpha$ determine the *constant* « spatial extension » of a such wave function, which may be taken to be $R' = R/\alpha$. As outlined above, $\alpha$ should be considered as a small « cut-off ». Such wave functions may indeed appear to be more correct realizations of a « pure momentum state » than a plane wave, because, *in contrast to plane waves:* (i) they obey Schrödinger's equation (except at the singularity $\mathbf{x} = \mathbf{a}(t)$ and on the « cancellation sphere » $|\mathbf{x} - \mathbf{a}(t)| = R/\alpha$), and (ii) they are square-integrable in space. In replacing $A_{R\alpha}$ by $R/r$, one gets momentum states that satisfy Schrödinger's equation everywhere except at the singularity and that vanish at spatial infinity, though without having a finite square integral. The fact that the evolution of the momentum is governed by Newton's second law seems to make the transition from classical mechanics to wave mechanics quite transparent: classical mechanics would envisage as a point particle what is in fact a spatially extended wave. And since the « pure momentum state » remains so only in a uniform force field, classical mechanics would be valid only in the case that the force fields vary slowly in space. We have thereby proved that, *contrary to the standard presentation of QM, a « quantum particle »* (as defined by a square-integrable wave function obeying Schrödinger's equation) *may have a finite and constant spatial extension* **and** *a well-defined momentum. Moreover, the amplitude of the wave function increases as 1/r towards a point singularity which moves according to Newton's second law.*[7] This is much in the line of de Broglie's ideas (see *e.g.* Ref. 17), though it is also compatible with Schrödinger's conception according to which the « matter waves » are the true reality.

A question arises then: how is this result compatible with Heisenberg's uncertainty principle? The answer is, obviously, that the latter applies (as a consequence of the QM postulates regarding the « collapse of the wave function ») to the outputs of *measurements* of momentum and position. Whereas the present result applies to the objective description of a « quantum particle » (wave function) in a spatially uniform force field, thus excluding a necessarily perturbating action of measurement. This result is hence compatible with the axioms of QM if they are taken in the operational sense, and it has no immediate consequence on the possible predictions that one can make in QM. We also note that, since the spatial extension of a « pure momentum » wave function is *arbitrary*, one still may say in a restricted sense that « the particle has not a well-defined position ». (This is no surprise if one considers that there is indeed no point particle but an extended wave!) Another one might consider, on the contrary, that « the position of the particle » is perfectly defined at any time, as the position $\mathbf{a}(t)$ of the singularity! The present result may contribute to give hope that one may go beyond QM towards a theory based on Schrödinger's notion of « real waves ». In any case, it indicates that QM as it is does not preclude Schrödinger's viewpoint (which appears essentially compatible with de Broglie's), at least at the stage of the scalar Schrödinger equation that does not integrate the notion of the quantum spin.

---

[6] Probably, $A$ must also have a singularity if one demands instead that $A$ vanish outside a bounded open domain D and be harmonic in D, but perhaps not on its boundary: *e.g.* this is true if D is the ball $|\mathbf{x}|<R$, as for $A_{R\alpha}$[7, p. 390].

[7] If one accepts a one-dimensional singularity, the behaviour near this singularity will have to be determined, *e.g.* it is not guaranteed in advance that a square-integrable function may be obtained. Anyhow, the amplitude will also undergo a pure translation with the uniform group velocity obeying Newton's second law (Eqs. (3.10, 3.14, 3.18)).

Finally, if one tries to interpret the wave function $\psi$ as a real wave, one may venture to guess which physical quantity could wave as $\psi$. If one dares to think of a mechanical wave in a fluid « micro-ether », it certainly could not be the displacement – due to the singularity. Similarly, it seems to be excluded that $\psi$ could be the pressure, because the micro-ether should be barotropic (Romani [13], Arminjon [1]), so an infinite amplitude would also occur for the density oscillation. Instead, the real part of $\psi$ could be perhaps a (any?) component of the velocity **u** of the flow, or instead, possibly, of **grad u**. This suggests again – but this time directly from the quantum-mechanical framework – the idea (tracing back at least to Kelvin, with *e.g.* Romani [13] and Winterberg [22-23] as recent defenders) that the elementary particles could be ideal vortices in a perfectly fluid « micro-ether ».

## 4. Questions for a (relativistic) quantum mechanics in a gravitational field

*4.1 General framework and the question of the reference frame and the reference time*

We adopt the general framework of theories in which gravitation changes the flat Galilean space-time metric into a curved metric $\gamma$. The latter is the physical metric, in the sense that, for example, the proper time defined along a trajectory as the integral of

$$d\tau = (\gamma_{\mu\nu}\, dx^\mu\, dx^\nu)^{1/2}/c, \qquad (4.1)$$

is assumed to be the time that may be measured by a physical clock moving along this trajectory. As it has been emphasized previously [3], this assumption of a « physical » pseudo-Riemannian metric does not imply *per se* that the relativity principle (principle of manifest covariance) must remain true in the presence of gravitation. In any such theory, including general relativity, the Poincaré-Einstein synchronization convention can be applied only to infinitesimally distant clocks [10]. At first sight, this seems to reinforce the relativity of simultaneity, because clocks may then be synchronized, in general, only along an open line in space. However, there exists (always locally *i.e.* in an open neighborhood of any space-time point, and indeed in the space-time as a whole, unless a « pathological » space-time such as Gödel's is considered), a « globally synchronized » reference frame (GSRF), *i.e.*, such that the $\gamma_{0i}$ components are zero in some adapted local coordinates $(x^\mu)$. (« Adapted » coordinates are ones in which any observer bound to the reference frame has constant space coordinates $(x^i)$.) The Poincaré-Einstein synchronization extends to the whole of a GSRF and defines, in this frame, a privileged family of time coordinates, that exchange by space-independent transformation of time: $x'^0 = \phi(x^0)$, under which the « $\gamma_{0i} = 0$ » condition is stable [3].

Now we wish to extend Schrödinger's wave mechanics to the case with a gravitational field. To this end, we just need to make sure that the tools which are used there remain at our disposal with a curved space-time metric $\gamma$. As noted in Section (2.1), the correspondence, at any given point $X$ of the extended configuration space, between the wave equation (*here of order $n \leq 2$*) and the dispersion equation demands that one restricts oneself to infinitesimally linear coordinate changes; and, for that purpose considered alone, the class of locally geodesic coordinate systems appears naturally. However, a clear space/time separation, as it occurs in Newtonian physics, has to be used so as to define the group velocity and the associated Hamiltonian system (2.11) (*cf.* Section (2.2)). Finally, the latter system is assumed to give the same set of trajectories as a certain classical Hamiltonian (2.12) (*cf.* Section (2.3)).

Therefore, we *a priori* need to introduce a preferred time coordinate (up to a scale factor) and hence to restrict ourselves to some subclass of coordinate systems, two of which should exchange by a purely spatial *and* infinitesimally linear transformation: restricting the discussion to a « one-particle » system, the admitted transformations are given by

$$x'^0 = ax^0, \quad x'^i = \phi^i(x^1, x^2, x^3),$$
$$(\partial^2 x'^i/\partial x^j \partial x^k)(X) = 0. \quad (4.2)$$

(In a *subsequent* step, one may examine the covariance of the obtained wave equation under a larger group of transformations. However, the need to fix the time coordinate will be further justified below: see after Eq. (4.4).) Furthermore, a classical (though « relativistic ») Hamiltonian system (2.12) must be found.[8] The first step is hence to ask: in the curved space-time of some theory of gravitation, can one select *in a natural way* a group of transformations of the form (4.2)? Note that such selection includes the choice of a particular reference frame.

In view of the outmost importance taken by the notion of a global simultaneity in Newtonian physics (which is the concrete arena for Hamiltonian systems), it indeed seems *natural* to require that the preferred time $t$ is a globally synchronized time, which means that a GSRF should be considered. However, there is a continuous class of GSRF 's, that undergo a « pure strain » (*i.e.*, no « rotation ») with respect to each other, and one would have to ask if one may then reexpress the wave mechanics covariantly from one to another. In what follows, *one* GSRF will be considered. The coordinate changes internal to such frame are [3]

$$x'^0 = \phi(x^0), \quad x'^i = \phi^i(x^1, x^2, x^3). \quad (4.3)$$

The further restriction to locally geodesic coordinate systems [$\gamma_{\mu\nu,\rho}(X) = 0$] is then impossible in general (no LGCS is in general bound to the given GSRF, in other words the local observer of the given frame is in general not in a « free fall »). One class of coordinate systems, stable under infinitesimally linear spatial changes, and compatible with (4.3), appears naturally: those systems whose space coordinates are locally geodesic at $X$ for the spatial metric **g** in the given GSRF:

$$g_{ij,k}(X) = 0 \quad (i, j, k = 1, 2, 3). \quad (4.4)$$

The intersection group (4.3)-(4.4) is still too large as compared with (4.2), because the time coordinate is not fixed. It is easy to verify, by an explicit calculation, that the operation of an arbitrary space-independent transformation of time, Eq. (4.3)$_1$, alters considerably the form of the wave equation (as well as that of the dispersion equation). *I.e.*, in order to state an exact form for the wave equation, it is really necessary that the time coordinate is not allowed to be changed by (4.3)$_1$. Moreover, the explicit form of the Hamiltonian $H$ itself (which is furnished by classical physics) is changed also, after a change $t = \phi(T)$, to

$$K(\mathbf{p}, \mathbf{x}, T) = \phi'(T) H(\mathbf{p}, \mathbf{x}, \phi(T)). \quad (4.5)$$

Thus, the fixing of the time coordinate is physically imposed. Actually, the author has proposed a preferred-frame, *scalar* theory of gravitation, according to which gravitation breaks the Lorentz invariance (see Refs. 1, 2, 4). The preferred frame (« macro-ether » or simply « ether ») of this theory is, by hypothese, a GSRF: a global simultaneity is defined in terms of the « absolute time » $T$, which is that measured by an observer bound to the ether and far enough from massive bodies. Hence, the answer to the above question is that, *in a preferred-frame theory of gravitation* as the one just evoked, one group (4.2) appears naturally, namely those coordinate systems whose time component is $x^0 = \alpha T$ with $T$ the absolute time, and whose spatial part verifies (4.4); and that it is *not* the case for theories starting from the relativity postulate (or here the principle of « manifest covariance »).

---

[8] Although it indeed must be relativistic in the sense of being consistent with a pseudo-Riemannian space-time, with the mass increase with velocity, etc., the system (2.12) must be expressed with such preferred time coordinate. That is, $H$ in Eq. (2.12) is a Hamiltonian with the *space*, not the space-time, as the configuration space (in the case of one particle); and with a true *time coordinate*, *not* the proper time $\tau$ of Eq. (4.1), as the time $t$, for the simple reason that $\tau$ is defined only once the line in space-time is known – whereas, in a « true » Hamiltonian system (2.12), the time $t$ is defined independently of the position in the configuration space!

*4.2 Comments on special relativity and on the covariance of Hamilton's equations*

(i) In the case of a *flat* space-time metric (special relativity), a particular class of GSRF 's does appear naturally, namely that of the (global) inertial frames (those frames which are defined by a Galilean coordinate system). And in any inertial frame, an even smaller group than (4.2) occurs naturally: that of the *global* Cartesian coordinates (combined with the change of the time scale *a*). In that case, one thus naturally builds one wave mechanics per inertial frame, and the question is then: what happens in the transition from one inertial frame to another? It seems that, as far as only the wave equation is concerned, nothing happens: if a wave equation is deduced, by the quantum correspondence (2.18), from a Lorentz-covariant expression of the Hamiltonian, $H = H(\mathbf{p},\mathbf{x}, t)$, then this wave equation itself is Lorentz-invariant, of course.

(ii) At first sight, it seems that the *a priori* restriction to coordinate changes defined by (4.2) is not imposed by the Hamiltonian dynamics *in itself*, but instead by the *correspondence* between the Hamiltonian and the wave equation. Indeed, a Hamiltonian system like (2.12) remains in the same form even after the most general canonical transformation of the extended phase space (although the Hamiltonian is changed if the transformation depends on time), and these transformations make a very large class. In such transformations, the position $\mathbf{x}$, momentum $\mathbf{p}$, and *time t* variables are considered on just the same footing: the only condition is that the 1-form $\mathbf{p}.d\mathbf{x} - H\,dt$ is invariant up to an exact 1-form $dS$ [5, p. 239]. However, the covariance of the Hamiltonian system (2.12) under canonical transformations is not in the same sense as (and indeed it is a weaker statement than), for instance, the covariance of the geodesic equation of motion for a free particle in a gravitational field (in general relativity and other « metric theories »). To be precise, consider a merely *spatial* transformation:

$$x'^0 = x^0, \qquad x'^i = \phi^i(x^1, x^2, x^3). \quad (4.6)$$

This may be seen as a canonical transformation with generating function

$$\Phi = \Sigma_j \, \phi^j(x^1, x^2, x^3) \, p'_j \quad (4.7)$$

By definition, the canonical variables transform indeed as

$$x'^i = \frac{\partial \Phi}{\partial p'_i} = \phi^i(x^1, x^2, x^3),$$
$$p_i = \frac{\partial \Phi}{\partial x^i} = \frac{\partial x'^j}{\partial x^i} p'_j \quad (4.8)$$

and the Hamiltonian system (2.12) is left exactly unchanged, only the new variables $x'^i$ and $p'_i$ are substituted for $x^i$ and $p_i$. However, if the covariance were as « rich » as in the case of the geodesic equation, then the two sides of Eq. $(2.12)_1$ would transform under (4.6) as a covector. It is obvious that the right-hand side of $(2.12)_1$: $-\partial H/\partial x^j$, does so. But the left-hand side becomes:

$$\frac{dp'_j}{dt} = \frac{d}{dt}\left(\frac{\partial x^i}{\partial x'^j} p_i\right) = \frac{\partial x^i}{\partial x'^j} \frac{dp_i}{dt} + \frac{\partial^2 x^i}{\partial x'^j \partial x'^k} \frac{dx'^k}{dt} p_i, \quad (4.9)$$

it is therefore a covector only for space transformations that are *infinitesimally linear* at the point considered, again. In other words, Eq. $(2.12)_1$ is covariant only if the variation of $\partial x^i/\partial x'^j$ along the trajectory is « forgotten » – in contrast to the geodesic equation, in the transformation of which this variation is accounted for, as one easily checks. This difficulty comes from the fact that, in Hamilton's equations, $(dp_i/dt)$ is indeed not a true covector but instead the *projection*, on the $dp_i$ part of the coordinate system $(dx^i, dp_i)$ on $T(T^*M)_{(\mathbf{x},\mathbf{p})}$ (deduced from the data of the coordinate system $x^i$ on M), of a tangent vector $\xi \in T(T^*M)_{(\mathbf{x},\mathbf{p})}$. Only in the case where M is *vector space* E is there a coordinate-independent decomposition of $\xi$ into a vector $d\mathbf{x}/dt$ and a covector $d\mathbf{p}/dt$ (due to the fact that, in that case, $T(T^*M)_{(\mathbf{x},\mathbf{p})}$ is canonically identified to $E \times E^*$). This canonical identification (implying an intrinsic decomposition for $\mathbf{x}$) is defined through the intermediate of the class of the *linear* coordinate systems (the linear bijections of E onto $\mathbf{R}^N$) (see Dieudonné [8, pp. 22-23]).

Thus, the *a priori* restriction to transformations all having the form (4.2) may also be seen as resulting from the need to fix the time coordinate (as it is fixed in physics that furnishes the classical Hamiltonian and that demands a precise form of the wave equation), plus the need that Hamilton's equations be *truly* covariant under the allowed coordinate transformations.

*4.3 Klein-Gordon equation in a gravitational field*

As an illustrative test of the foregoing framework, let us investigate the possible extension of the free Klein-Gordon equation (3.6) to the case where a gravitational field is present. Thus, we consider a massive test particle subjected only to the gravitation. We first have to ask whether a Hamiltonian $H$ does govern the motion of a such « free » test particle in a gravitational field. This feature may depend on the law of motion in the theory of gravitation. Precisely, it may depend on whether the test particle follows a geodesic line of metric $\gamma$ (as in « metric theories ») or, instead, its motion is defined by an extension of Newton's second law with a gravity acceleration field $\mathbf{g}$, generally incompatible with geodesic motion (as in the « ether theory of gravitation » (ETG) studied by the author). The conclusion of § 4.1, that the necessary restriction on the coordinate systems (Eq. (4.2)) may be fulfilled in a natural way only in such preferred-frame theory, gives some justification for starting the present discussion with the investigated ETG. However, the result that will be obtained, and that applies to a static gravitational field, is also true in metric theories since, in the case of a static field, the equation of motion in the ETG is the geodesic equation, thus the same as in metric theories [2-3].

In the ETG, the energy of the particle in the gravitational field is defined by

$$e \equiv \beta E = \beta m c^2 \gamma_v, \qquad (4.10)$$

where $\beta \equiv \sqrt{\gamma_{00}}$, $\gamma_v \equiv 1/\sqrt{(1-v^2/c^2)}$, the velocity vector $\mathbf{v}$ (in the preferred frame) and its modulus $v$ being evaluated in terms of the local physical standards of space and time:

$$v^i \equiv dx^i/dt_\mathbf{x} = (1/\beta)u^i, \qquad u^i \equiv dx^i/dT,$$
$$v \equiv [\mathbf{g}(\mathbf{v},\mathbf{v})]^{1/2} \equiv (g_{ij} v^i v^j)^{1/2}. \qquad (4.11)$$

The energy defined for metric theories by Landau & Lifchitz [10], turns out to have the same general expression (4.10). It is conserved if and only if the field is constant: this is true for the ETG as well as for metric theories [3]. It is hence natural to try $e$ as a Hamiltonian $H(\mathbf{p},\mathbf{x},T)$, with a canonical momentum $\mathbf{p}$ that is yet to be determined. The Lagrangian associated with $H$ is defined by Legendre transformation:

$$L(\mathbf{x},\mathbf{u},T) = p_j(\mathbf{x},\mathbf{u},T) u^j - e, \quad p_j = \partial L/\partial u^j. \quad (4.12)$$

Differentiating $(4.12)_1$ with respect to $u^i$ and using $(4.12)_2$, one gets

$$u^j \partial p_j/\partial u^i = \partial e/\partial u^i = m \gamma_v^3 g_{ij} u^j/\beta. \quad (4.13)$$

This suggests trying $\partial p_j/\partial u^i = m\gamma_v^3 g_{ij}/\beta$, which gives

$$p_j = \int_0^1 \frac{\partial p_j}{\partial u^i}(\xi \mathbf{u}) u^i d\xi = m g_{ij} v^i \int_0^1 \frac{d\xi}{\left(1-\xi^2 v^2/c^2\right)^{3/2}}$$
$$= m \gamma_v g_{ij} v^i. \qquad (4.14)$$

Hence, *the canonical momentum is the usual momentum*. The Lagrangian $(4.12)_1$ is then found to be

$$L = -mc^2 \beta \sqrt{1 - v^2/c^2}. \qquad (4.15)$$

Inserting this $L$ in $(4.12)_2$, one verifies that $p_j$ is given by Eq.(4.14), and from $L$ the Legendre transform gives indeed $e$, Eq. (4.10). In covector form, the expression of Newton's second law in the ETG [3] gives:

$$m\gamma_v\left(-c^2 \frac{\beta_{,i}}{\beta}\right) = \frac{Dp_i}{Dt_{\mathbf{x}}} \equiv \frac{D_0 p_i}{Dt_{\mathbf{x}}} - \frac{1}{2} \frac{\partial g_{ij}}{\partial t_{\mathbf{x}}} p^j \qquad (4.16)$$

where

$$p^j \equiv g^{jk} p_k \quad \text{and} \quad \frac{D_0 p_i}{Dt_{\mathbf{x}}} = \frac{dp_i}{dt_{\mathbf{x}}} - \frac{1}{2} g_{jk,i} \frac{dx^j}{dt_{\mathbf{x}}} p^k,$$

and where

$$\frac{\partial}{\partial t_{\mathbf{x}}} \equiv \frac{1}{\beta} \frac{\partial}{\partial t} \equiv \frac{c}{\sqrt{\gamma_{00}}} \frac{\partial}{\partial x^0}. \qquad (4.17)$$

Using (4.16), one finds after some algebra:

$$\frac{d}{dT}\left(\frac{\partial L}{\partial u^i}\right) - \frac{\partial L}{\partial x^i} = \frac{1}{2} \frac{\partial g_{ij}}{\partial T} p^j. \qquad (4.18)$$

This proves that the equations of motion in the ETG derive from the Hamiltonian $e$, Eq. (4.10) (which will be reexpressed below as a function $H$ of the conjugated variables $\mathbf{p}$ and $\mathbf{x}$), if and only if the spatial metric in the preferred frame, $\mathbf{g}$, does not depend on time. The specific form of the metric assumed in the ETG is *not* used in deriving (4.18). If one considers the case of any *static* metric $\gamma$ in a metric theory, then, in the (unique) reference frame where the metric is static, the equations of motion are just the same as in the ETG [3] and so *the same result applies*. Note that the time coordinate is fixed up to a constant factor by the assumption that $\gamma$ is static [3]. In the ETG, the reference frame where the metric is static is automatically the preferred frame; moreover, the spatial metric $\mathbf{g}$ can be time-independent only if the scalar gravitational field of the ETG is itself time-independent. (This scalar field is just $\beta \equiv \sqrt{\gamma_{00}}$, expressed with $x^0 = cT$.)

Now the purely material energy $E = mc^2 \gamma_v$ verifies the well-known relation

$$E^2 - \mathbf{p}^2 c^2 = m^2 c^4, \qquad (4.19)$$
$$\mathbf{p}^2 \equiv \mathbf{g}(\mathbf{p}, \mathbf{p}) \equiv g^{ij} p_i p_j,$$

whence

$$[H(\mathbf{p}, \mathbf{x})]^2 = \beta(\mathbf{x})^2 c^2 g^{ij}(\mathbf{x}) p_i p_j + \beta(\mathbf{x})^2 m^2 c^4. \qquad (4.20)$$

The correspondence (2.18) gives, unambiguously as noted in part 2, the wave equation

$$-\hbar^2 \frac{\partial^2 \psi}{\partial T^2} = -\hbar^2 \beta^2 c^2 g^{ij} \frac{\partial^2 \psi}{\partial x^i \partial x^j} + \beta^2 m^2 c^4 \psi, \qquad (4.21)$$

which is valid in any coordinates verifying (4.4). (Note that, had we not the arguments of part 2 to « put $\mathbf{x}$ before $\mathbf{k}$ » in each term of the dispersion equation, the correspondence (2.18) would be highly ambiguous.) Among those, we may select ones in which $g^{ij} = g_{ij} = \delta_{ij}$. We then recognize that Eq. (4.21) may be rewritten in any spatial coordinates as

$$\Delta_{\mathbf{g}} \psi - \frac{1}{c^2 \beta^2} \frac{\partial^2 \psi}{\partial T^2} - \frac{1}{\lambda^2} \psi = 0, \qquad (4.22)$$
$$\Delta_{\mathbf{g}} \psi \equiv \frac{1}{\sqrt{g}} \frac{\partial}{\partial x^i}\left(\sqrt{g}\, g^{ij} \frac{\partial \psi}{\partial x^j}\right),$$

where $g \equiv \det(g_{ij})$. Since the metric $\gamma$, hence $\beta \equiv \sqrt{\gamma_{00}}$, has been assumed time-independent, Eq. (4.22) may still be put in the form

$$\Delta_{\mathbf{g}} \psi - \frac{1}{c^2} \frac{\partial^2 \psi}{\partial t_{\mathbf{x}}^2} - \frac{1}{\lambda^2} \psi = 0. \qquad (4.23)$$

Equation (4.23) is coordinate-independent, but it must be understood that metric $\mathbf{g}$ is the spatial metric in the unique reference frame where the space-time metric $\gamma$ is static, and that $t_{\mathbf{x}}$ is the local time in this same reference frame. (The definition (4.17) of the derivative with respect to $t_{\mathbf{x}}$ is convenient, but it is valid only in coordinates adapted to this frame.) Note that Eq. (4.23) makes sense also for the case where the metric $\gamma$ is not static, although one has then one equation per reference frame selected. Thus, it is *not* a manifestly covariant equation. It is interesting to note that the operator $\Delta_\mathbf{g} - (1/c^2)\partial^2/\partial t_\mathbf{x}^2$ is the same as in the field equation for the scalar gravitational field in the ETG [1, 2, 4].

The manifestly covariant generalization of the Klein-Gordon equation is obvious:

$$\bullet_\gamma \psi + \frac{1}{\lambda^2}\psi = 0, \qquad (4.24)$$

$$\bullet_\gamma \psi \equiv (\psi^{;\mu})_{;\mu} = \frac{1}{\sqrt{-\gamma}}\frac{\partial}{\partial x^\mu}\left(\sqrt{-\gamma}\,\gamma^{\mu\nu}\frac{\partial \psi}{\partial x^\nu}\right),$$

where $\gamma \equiv \det(\gamma_{\mu\nu})$. This extension is used in works on quantum field theories in curved space-time, *e.g.* Wald [20]. However, it is not equivalent to Eq. (4.23), *even for a static metric* $\gamma$. Indeed, let us assume that the frame F, in which Eq. (4.23) is written, is globally synchronized, and let us select coordinates bound to F, such that $\gamma_{0i} = 0$, and such that, at the space-time point $X$ considered, one has moreover: (i) $g_{ij,k} = 0$, (ii) $g^{ij} = g_{ij} = \delta_{ij}$, (iii) $\gamma_{00} = 1$ and (iv) $\gamma_{00,0} = 0$. (Such choice is always possible with an internal coordinate change (4.3).) In such coordinates, one has for a general metric $\gamma$:

$$\bullet_\gamma \psi \qquad (4.25)$$
$$= -\left(\Delta_\mathbf{g}\psi - \frac{1}{c^2}\frac{\partial^2\psi}{\partial t_\mathbf{x}^2}\right) + \left(\psi_{,0}\,g_{,0} + \psi_{,j}\gamma_{00,j}\right)/2.$$

Due to the last term in the r.h.s. of (4.25), Eqs. (4.23) and (4.24) are not equivalent if the metric is static.

In summary, the « quantum » correspondence (2.18), based on the relation (2.14) between the classical Hamiltonian and the dispersion equation of the wave operator, works perfectly well in the case of a particle moving « freely » in a static gravitational field. According to the principles of *wave mechanics*, the wave equation (4.23) thus obtained is the *unique* extension of the free Klein-Gordon equation to a static gravitational field – and it makes sense also for a general gravitational field. *But* this extension *differs*, already for a static field, from the obvious extension of the K-G equation according to the principle of *manifest covariance*, Eq. (4.24). As a matter of fact, the wave-mechanical extension (4.23) is explicitly frame-dependent. This means that a preferred reference frame must exist, in order that wave mechanics can make sense in the presence of gravitation: in the idealized case of a static gravitational field, it is natural to assume that the preferred frame is the reference frame where the field is static. But if one wishes to extend wave mechanics to a *general* gravitational field, then one should postulate Eq. (4.23), and this means really that a preferred reference frame must exist in the general case. Another argument goes as follows: if the « wave-mechanically correct » extension of the K-G equation could be manifestly covariant, then it would be Eq. (4.24), and this cannot be correct since Eq. (4.24) does not coincide with Eq. (4.23) in the static case, for which the wave-mechanically correct extension of the K-G equation is undoubtedly Eq. (4.23). The result obtained for the free K-G equation makes it plausible that *the correct extension of any quantum-mechanical equation should be a preferred-frame equation, obtained simply by substituting the spatial metric $\mathbf{g}$ in the preferred frame for the Euclidean metric, and the local time in the preferred frame, $t_\mathbf{x}$, for the inertial time.* This hypothese should be checked by its consequences, of course.

However, the « wave-mechanically correct » extension (4.23) of the K-G equation does *not* satisfy the equation for continuum dynamics in metric theories of gravitation, which is $T^{\mu\nu}{}_{;\nu} = 0$. The energy-momentum tensor **T** of the K-G field is defined by [6]:

$$T_{\mu\nu} \equiv \psi^*{}_{,\mu}\,\psi_{,\nu} + \psi_{,\mu}\,\psi^*{}_{,\nu} - \Lambda\gamma_{\mu\nu},$$

$$\Lambda \equiv \psi^*{}_{,\mu}\,\psi^{,\mu} - \frac{|\psi|^2}{\lambda^2}. \qquad (4.26)$$

(One may verify that this tensor **T** is the symmetrical energy-momentum tensor canonically associated, by the standard procedure of Ref. 10, §94, with the Lagrangian *L*.) Using locally geodesic coordinates, it is easy to verify the following generally-covariant identity:

$$T_\mu{}^\nu{}_{;\nu} = \psi_{,\mu} \, (\bullet_\gamma \psi + \psi/\lambda^2)^* + \text{conjugate}. \quad (4.27)$$

Combining Eqs. (4.25) and (4.27), one finds that, if $\psi$ verifies the *wave-mechanically correct* extension of the K-G equation, Eq. (4.23), then $T^{\mu\nu}{}_{;\nu} \neq 0$, even in the static case. (In the static case, the dynamical equation is $T^{\mu\nu}{}_{;\nu} = 0$ also in the « ether theory of gravitation » [4].) In contrast, Eq. (4.27) shows immediately that the *generally-covariant* extension of the K-G equation, Eq. (4.24), does obey $T^{\mu\nu}{}_{;\nu} = 0$.

*4.4 Stationary states of the gravitational K-G equation*

Obviously, it is first for the case of a constant gravitational field that one may expect that stationary solutions of the wave equation might have similar properties as in the absence of gravitation and, in particular, allow to define relevant « energy levels ». In the case of a static gravitational field, the stationarity (Eq. (3.2)) should be expressed with the preferred time *t* (up to a constant factor) in terms of which the metric is static. In a more general case, we again encounter the problem of the ambiguity in the time coordinate: a preferred time *t* must exist, in order that Eq. (3.2) can make sense. Inserting (3.2) into the wave equation (4.23), we get instead of Eq. (3.6):

$$\phi \, (\Delta_{\mathbf{g}_t} a - a/\lambda^2) = \frac{a}{c^2 \beta^2} \left( \phi'' - \frac{\beta_{,t}}{\beta} \phi' \right), \quad (4.28)$$

in which the Laplace operator (Eq. (4.22)) corresponds to the usually time-dependent Riemannian metric $\mathbf{g}_t$. Hence, in the case of a *static* gravitational field, Eq. (4.28) implies $\phi''/\phi$ = Const. $\equiv -\omega^2$ and we get a spatial eigenvalue problem:

$$\beta^2 \, [a/\lambda^2 - \Delta_{\mathbf{g}} a] = (\omega^2/c^2) \, a. \quad (4.29)$$

*I.e.*, in the static case, any stationary solution of the gravitational K-G equation corresponds to an eigenfunction of a spatial operator; and the difference between two eigenfrequencies may be interpreted as the frequency of absorption/emission of an electromagnetic radiation. However, in the general case, a stationary solution (3.2) does not seem to correspond to an eigenfunction of a purely spatial operator.

**5. Concluding remarks, including a conjecture**

In this paper, it has been attempted to give rigorous results although, for the purpose of simplicity, a rather old-fashioned mathematical language has been mainly used – namely the language of the coordinates and their changes. Thus, except for some allusions, it has not been much insisted on the heuristic concept of ether that guides the author's work. According to this concept, principally due to Romani [13], all matter and *non-gravitational* fields are just flows in a perfectly fluid « micro-ether » (named simply ether by Romani). Romani proposed a definite structure of torus vortex for the electron, and tried to explain all other known particles as combinations (or « complexes ») of torus vortices. The electromagnetic and nuclear fields were considered as particular expressions of the essentially unique force field, which would be the (microscopic) field of ether pressure. Romani may have gone too far in trying to assign such definite structure to the elementary particles and fields, before having a consistent theoretical framework at his disposal. However, Romani's general idea has a heuristic value. The results of Waite [19] support the essential idea that the « quantum » (discrete) aspect in microphysics might come from the individuality enjoyed by a special flow topology such as a closed vortex. (But even such special flow is never fully distinct from the remainder of the fluid, which could explain the basic unseparability of quantum theory. Also note that a *point* singularity gives an identifiable discrete character as well.) However, an important difference is that Waite considers that the electromagnetic force could be elementary, constitutive of the substratum; whereas Romani assumed that the property of an electric charge makes sense only for an entire vortex, which is more obviously consistent with the quantization of the charge.

A serious lack in Romani's model is that there is no link between the ether and the inertial frames. In the theory of gravitation studied in Refs. 1, 2, 4, the preferred reference frame or « macro-ether » is defined by the average motion of the « micro-ether ». As to the gravitation force, it would be due [1] to the *macroscopic* part of the ether pressure gradient, more exactly the macroscopic ether pressure would *decrease* towards the gravitational attraction, hence towards the concentration of matter. This assumption is consistent with the notion that material particles could be complexes of vortices, because the pressure is indeed lower in the vicinity of a vortex. In any case, it leads to a consistent scalar theory of gravitation. Thus, gravitation, contrary to the other interactions, would depend only on the macroscopic effect of the imagined microscopic flows. So the very definition of the gravitational force has a macroscopic nature, involving a statistics over a huge number of micro-objects, and it does not make sense to isolate the contribution of one microscopic (quantum) object to the gravitational force – although the gravitational force does have to be taken into account for an exact dynamics of even a small number of microscopic objects.

In summary, the heuristic concept of ether envisaged here sees matter and non-gravitational fields as microscopic flows in a perfectly fluid substratum, the necessity of quantization being due to the discrete distribution of flow singularities (vortices). On the other hand, gravitation would be a continuous variation in the macroscopic ether pressure, due to the accumulation of a great number of vortices. If this view is correct, then an important conclusion follows: *gravitation has not to be quantized*. The question is then to build a consistent theory involving quantum mechanics in the curved space-time of a non-quantized gravitation theory. A framework for writing QM in a curved space-time, consistent with what is thought to be the essence of Schrödinger's wave mechanics, has been presented here. The main feature of this consistent extension is that it needs that a physically preferred reference frame must exist in the presence of gravitation – as is true for the theory investigated in Refs. 1, 2, 4. The problem is then to define the source of the classical gravitational field. We assume that the source of the gravitational field is simply the macroscopic, *classical* energy-momentum tensor, **T** (more precisely, in the theory [1-2, 4], only the $T^{00}$ component, in coordinates bound to the preferred frame, and with $x^0 = cT$ where $T$ is the absolute time: see the end of § 4.1). This assumption presupposes that the question of the transition from quantum theory to classical, macroscopic theory is solved, of course. However, this is a different question from the question whether gravitation must be quantized or not. Moreover, the former question is more of a philosophical character, for in fact **T** is defined phenomenologically. Hence, the author definitely conjectures that there is no need for a quantum gravity.

However, it has been found a conflict between the « wave-mechanically correct » gravitational extension of a wave equation of QM and the dynamical equation in theories of gravitation in a curved space-time (see the end of § 4.3). This is considered to be a serious problem. *A priori*, we do not expect that this problem could be consistently solved by modifying the definition of the non-gravitational energy-momentum tensor **T** in the presence of a gravitational field. Indeed, it seems that, on the contrary, the definition of tensor **T** should be the same whether there is a gravitational field or not. If the problem cannot be solved otherwise, it might lead one to try to modify QM in such a way that it becomes compatible with dynamics in a gravitational field. This, of course, would be an ambitious task!